\documentclass[a4paper,10pt]{article}
\usepackage{
graphicx,
hyperref,
amsmath,
amssymb,
charter,
xcolor,
ifluatex,
longtable,
booktabs,
multirow,
bbold,cancel}

\usepackage{
graphicx,
hyperref,
amsmath,
amssymb,
charter,
xcolor,
ifluatex,
longtable,
booktabs,
multirow,
bbold}
\usepackage{colortbl}

\definecolor{Gray}{gray}{0.95}
\definecolor{RGray}{gray}{0.85}
\definecolor{CGray}{gray}{0.92}

\usepackage{multicol}
\definecolor{tit}{rgb}{0.1,0.2,0.4}
\definecolor{blus}{cmyk}{1,1,0,0.6}
\definecolor{verde}{cmyk}{0.92,0,0.59,0.25}

\usepackage{tikz}
\usetikzlibrary{matrix}

\usepackage{devanagari}
\usepackage{tipa}
\usepackage{amsmath,amssymb,amsfonts, bm}
\usepackage{dsfont}
\usepackage{epsfig}
\usepackage{graphicx}
\usepackage{slashed}
\usepackage{color}

\usepackage{caption}
\usepackage{subcaption}
\usepackage{slashed} 

\usepackage{adjustbox}
\usepackage{commath}
\usepackage{calc}

\usepackage{sidecap} 

\newcommand{\be}{\begin{equation}}
\newcommand{\ee}{\end{equation}}

\newcommand{\bea}{\begin{eqnarray}}
\newcommand{\eea}{\end{eqnarray}}

\newcommand{\bfig}{\begin{figure}}
\newcommand{\efig}{\end{figure}}

\usepackage{slashed}

\usepackage{commath}
\usepackage{calc}

\newcommand{\D}{{\cal D}}
\newcommand{\U}{{\cal U}}

\newcommand{\M}{{\cal M}}

\makeatletter
\newcommand*{\rom}[1]{\expandafter\@slowromancap\romannumeral #1@}
\makeatother

\usepackage{color,hyperref}
\definecolor{darkblue}{rgb}{0.0,0.0,0.3}
\hypersetup{colorlinks,breaklinks,
            linkcolor=darkblue,urlcolor=darkblue,
            anchorcolor=darkblue,citecolor=darkblue}

\def\1by3{\ensuremath{\frac{1}{3}}}
\def\4by3{\ensuremath{\frac{4}{3}}}
\def\2by3{\ensuremath{\frac{2}{3}}}

\providecommand*\url[1]{\href{#1}{#1}}
\renewcommand*\url[1]{\href{#1}{\texttt{#1}}}

\begin{document} 
%\maketitle
%\flushbottom

\allowdisplaybreaks
\vspace*{-2.5cm}
\begin{flushright}
{\small
IIT-BHU
}
\end{flushright}

\vspace{2cm}

\begin{center}
{\LARGE \bf \color{tit} Dark-technicolour at low scale }\\[1cm]

{\large\bf Gauhar Abbas$^{a}$\footnote{email: gauhar.phy@iitbhu.ac.in}   },~
{\large\bf Neelam Singh$^{a}$\footnote{email: neelamsingh.rs.phy19@itbhu.ac.in }   }  
\\[7mm]
{\it $^a$ } {\em Department of Physics, Indian Institute of Technology (BHU), Varanasi 221005, India}\\[3mm]

\vspace{1cm}
{\large\bf\color{blus} Abstract}
\begin{quote}
We discuss a low-scale realization of the dark-technicolour paradigm, where  the dark-technicolour scale is close to the electroweak scale.  This scenario provides an ultraviolet completion of the standard HVM, and predicts a dark-Higgs with mass $ m_{\rm DH } = 95.4$ GeV.  Moreover, the grand-unification scale in this framework can be as low as $1.18 \times 10^8$ GeV.

\end{quote}

\thispagestyle{empty}
\end{center}

\begin{quote}
{\large\noindent\color{blus} 
}

\end{quote}

\newpage
\setcounter{footnote}{0}

\section{Introduction}
\label{intro}
In 2017, in an interview to the CERN Courier, Weinberg asked the question,  ``Why is there something rather than nothing?"\cite{cern_courier}.  This question was a reference to the origin of the mass,  or the  flavour problem,  in the standard model.  There are several possible explanations to this question \cite{Abbas:2017vws}-\cite{Higaki:2019ojq}.  

One of the atypical explanations to the flavour problem arises from the VEVs hierarchy of the gauge singlet scalar fields  \cite{Abbas:2017vws,Abbas:2020frs,Abbas:2023dpf}.   This framework is referred  to as hierarchical VEVs model (HVM).   Origin of the  hierarchical VEVs lies in a dark-technicolour (DTC) dynamics, where these VEVs are sequential chiral multi-fermion condensates \cite{Abbas:2020frs}.   It turns out that a particular form  of HVM,  referred  to as standard HVM (SHVM)  \cite{Abbas:2023dpf},  can predict  the leptonic mixing parameters very precisely  in terms of the Cabibbo angle and masses of strange and charm quarks  along with the so-called flavonic dark-matter \cite{Abbas:2023ion}.   For a recent review on  HVM  and the flavour problem,   see  reference \cite{Abbas:2023ivi}. 

Technicolour (TC) gauge theories provide a natural origin of the electroweak symmetry breaking through the dynamical symmetry breaking  \cite{Weinberg:1975gm,Susskind:1978ms}.   QCD-like technicolour theories suffer from large flavour changing neutral currents (FCNC),  and are incompatible with electroweak precision measurement.  The presence of large FCNC pushes the extended technicolour (ETC) scale to approximately $10^6$ GeV,  which results in unrealistic fermion masses \cite{Abbas:2020frs}.    This problem is solved by non-QCD-like models, where the running coupling is absent \cite{walking1}-\cite{walking7}.

The DTC paradigm does not deviate from the QCD-like behaviour,  and solves the problem of FCNC and fermionic masses by decoupling them.  This means that the ETC scale remains  high ($> 10^6 $ GeV),  and fermionic masses originate from the DTC dynamics   \cite{Abbas:2020frs}.   The scale of the DTC dynamics,  $\Lambda_{\rm DTC}$,  is extremely high  (approximately $10^{12}$ GeV) in reference \cite{Abbas:2020frs}.   In this work,  we discuss a low-energy realization of the DTC dynamics,  which can be probed in experiments such as the Large Hadron Collider (LHC).

\section{Standard HVM}
We first briefly review SHVM \cite{Abbas:2023dpf}. In SHVM,  the charged fermion  masses  are obtained by the hierarchical VEVs of the gauge singlet fields $\chi_i$ after imposing a generic   $\mathcal{Z}_{\rm N} \times \mathcal{Z}_{\rm M} \times \mathcal{Z}_{\rm P}$ flavour symmetry on the SM \cite{Abbas:2023dpf}.  For instance, the $\mathcal{Z}_2 \times \mathcal{Z}_4 \times \mathcal{Z}_{14} $ symmetry is imposed on the SM and gauge singlet scalar fields $\chi_i$, as given in table \ref{tab1}.

\begin{table}[ht]
\begin{center}
\resizebox{\textwidth}{!}{
\begin{tabular}{|c|c|c|c||c|c|c|c||c|c|c|c||c|c|c|c|c|}
  \hline
  Fields                               &   $\mathcal{Z}_2$  &  $\mathcal{Z}_4$   &  $\mathcal{Z}_{14}$   & Fields   &  $\mathcal{Z}_2$   &  $\mathcal{Z}_4$ &  $\mathcal{Z}_{14}$ & Fields   & $\mathcal{Z}_2$  & $\mathcal{Z}_4$  &  $\mathcal{Z}_{14}$  & Fields  &  $\mathcal{Z}_2$   &  $\mathcal{Z}_4$  &  $\mathcal{Z}_{14}$ \\
  \hline
 $u_{R}$                        &     -   &     $ 1$          &    $\omega^{11}$ & $d_{R} $, $ s_{R}$, $b_{R}$    &     +    & $ 1$        &    $\omega^{12}$  & $ \psi_{L_3}^{q} $       &    +     &  $ 1$    &   $\omega^2$   &  $\tau_R$      &   +  &  $\omega^3$      &     $\omega$             \\
  $c_{R}$                       &     +   &    $ 1$          &    $\omega^6$   &  $\chi _4$                         &      +  &  $ 1$      &  $\omega^{13}$  &  $ \psi_{L_1}^\ell $                          &     +   & $\omega^3$     &  $\omega^{12}$                          &   $\nu_{e_R}$   &    +   &   $\omega$     &      $\omega^{8}$        \\
   $t_{R}$                        &     +   &    $1$         &    $\omega^{4}$   & $\chi _5$                         &      +  &  $1$    &  $\omega^{11}$  &  $ \psi_{L_2}^{\ell} $     &      +  & $\omega^3$      &  $\omega^{10}$     & $\nu_{\mu_R}$                   &     -  &   $\omega$    &   $\omega^3$                    \\
  $\chi _1$                        &      -  &   $ 1 $      &    $\omega^2$    &   $\chi _6$                          &      +  &  $ 1$      &   $ \omega^{10}$    &   $ \psi_{L_3}^{ \ell} $       &    +     &  $\omega^3$    &   $\omega^6$                                 & $\nu_{\tau_R}$                    &     -   &  $\omega$     &   $\omega^3$          \\
  $\chi _2$                   & +     &       $ 1$      &  $\omega^5$   & $ \psi_{L_1}^q $                          &      +  &  $1$      &  $\omega^{13}$  & $e_R$    &      -   &   $\omega^3$       &    $\omega^{10}$      &  $\chi_7 $                          &      -   &  $\omega^2$   &     $\omega^8$                                               \\
   $\chi _3$                  &    +   &       $ 1$    & $ \omega^2$        & $ \psi_{L_2}^{q} $     &      +  & $1 $      &  $\omega$  &   $ \mu_R$     &   +  & $\omega^3$       &     $\omega^{13}$      &  $ \varphi $                           &      +  &1     &   1                                            \\
  \hline
     \end{tabular}}
\end{center}
\caption{The charges of left- and right-handed fermions  and  scalar fields under $\mathcal{Z}_2$, $\mathcal{Z}_4$,  and $\mathcal{Z}_{14}$ symmetries for the normal mass ordering. The $\omega$ is the 4th and 14th root of unity corresponding to the symmetries $\mathcal{Z}_4$  and $\mathcal{Z}_{14}$,  respectively.}
 \label{tab1}
\end{table} 

The masses of the charged-fermions are  obtained by the  dimension-5 operators   given as,
\bea
\label{mass2}
{\mathcal{L}} &=& \dfrac{1}{\Lambda_{\rm F} }\Bigl[  y_{ij}^u  \bar{\psi}_{L_i}^{q}  \tilde{\varphi} \psi_{R_i}^{u}   \chi _i +     
   y_{ij}^d  \bar{\psi}_{L_i}^{q}   \varphi \psi_{R_i}^{d}  \chi _{i}   +   y_{ij}^\ell  \bar{\psi}_{L_i}^{\ell}   \varphi \psi_{R_i}^{\ell}  \chi _{i} \Bigr]  
+  {\rm H.c.},
\eea
$i$ and $j$   are family indices,  $ \psi_{L}^q,  \psi_{L}^\ell  $ denote  the  quark and leptonic doublets,  $ \psi_{R}^u,  \psi_{R}^d, \psi_{R}^\ell$ are the right-handed up,  down-type  quarks and  leptons,  $\varphi$ and $ \tilde{\varphi}= -i \sigma_2 \varphi^* $  are the SM Higgs field and its conjugate, and $\sigma_2$ is  the second Pauli matrix. 

The fermionic mass hierarchy is  explained in terms of  the   VEVs pattern    $ \langle \chi _4 \rangle > \langle \chi _1 \rangle $, $ \langle \chi _2 \rangle >> \langle \chi _5 \rangle $, $ \langle \chi _3 \rangle >> \langle \chi _6 \rangle $, $ \langle \chi _{3} \rangle >> \langle \chi _{2} \rangle >> \langle \chi _{1} \rangle $, and  $ \langle \chi _6 \rangle >> \langle \chi _5 \rangle >> \langle \chi _4 \rangle $.  The mass matrices of up,  down quarks and leptons read as,
\begin{align}
\label{mUD}
\M_\U & =   \dfrac{ v }{\sqrt{2}} 
\begin{pmatrix}
y_{11}^u  \epsilon_1 &  0  & y_{13}^u  \epsilon_2    \\
0    & y_{22}^u \epsilon_2  & y_{23}^u  \epsilon_5   \\
0   &  y_{32}^u  \epsilon_6    &  y_{33}^u  \epsilon_3 
\end{pmatrix},  
\M_\D = \dfrac{ v }{\sqrt{2}} 
 \begin{pmatrix}
  y_{11}^d \epsilon_4 &    y_{12}^d \epsilon_4 &  y_{13}^d \epsilon_4 \\
  y_{21}^d \epsilon_5 &     y_{22}^d \epsilon_5 &   y_{23}^d \epsilon_5\\
    y_{31}^d \epsilon_6 &     y_{32}^d \epsilon_6  &   y_{33}^d \epsilon_6\\
\end{pmatrix}, \\ \nonumber 
\M_\ell & =\dfrac{ v }{\sqrt{2}} 
  \begin{pmatrix}
  y_{11}^\ell \epsilon_1 &    y_{12}^\ell \epsilon_4  &   y_{13}^\ell \epsilon_5 \\
 0 &    y_{22}^\ell \epsilon_5 &   y_{23}^\ell \epsilon_2\\
   0  &    0  &   y_{33}^\ell \epsilon_2 \\
\end{pmatrix},
\end{align} 
where $\epsilon_i = \dfrac{\langle \chi _{i} \rangle }{\Lambda_{\rm F}}$ and  $\epsilon_i<1$.  

\begin{eqnarray}
\label{mass1b}
m_t  &=& \ \left|y^u_{33} \right| \epsilon_3 v/\sqrt{2}, ~
m_c  = \   |y^u_{22} \epsilon_2  - \dfrac{y_{23}^u  y_{32}^u \epsilon_5 \epsilon_6 }{y_{33}^u \epsilon_3} |  v /\sqrt{2} ,~
m_u  =  |y_{11}^u  |\,  \epsilon_1 v /\sqrt{2},\nonumber \\
m_b  &\approx& \ |y^d_{33}| \epsilon_6 v/\sqrt{2}, 
m_s  \approx \   |y^d_{22} - \dfrac{y_{23}^d  y_{32}^d}{y_{33}^d} | \epsilon_5 v /\sqrt{2} ,\nonumber \\
m_d  &\approx&  |y_{11}^d - {y_{12}^d y_{21}^d \over  y^d_{22} - \dfrac{y_{23}^d  y_{32}^d}{y_{33}^d} }   -
{{y_{13}^d (y_{31}^d y_{22}^d - y_{21}^d y_{32}^d )-y_{31}^d  y_{12}^d  y_{23}^d } \over 
{ (y^d_{22} - \dfrac{y_{23}^d  y_{32}^d}{y_{33}^d})  y^d_{33}}}   |\,  \epsilon_4 v /\sqrt{2},\nonumber \\
m_\tau  &\approx& \ |y^\ell_{33}| \epsilon_2 v/\sqrt{2}, ~
m_\mu  \approx \   |y^\ell_{22} | \epsilon_5 v /\sqrt{2} ,~
m_e  =  |y_{11}^\ell   |\,  \epsilon_1 v /\sqrt{2}.
\end {eqnarray}

The quark mixing angles are,
\begin{eqnarray}
\sin \theta_{12}  & \simeq &  { \epsilon_4 \over \epsilon_5}, ~
\sin \theta_{23}  \simeq   \sin \theta_{12}^{ 2},~
\sin \theta_{13}  \simeq   \sin \theta_{12}^{ 3} - 2 \frac{m_c}{m_t}.
\end{eqnarray} 

The neutrino masses are obtained  by adding   three right-handed  neutrinos $\nu_{eR}$, $\nu_{\mu R}$, $\nu_{\tau R}$  to the SM, and by writing the dimension-6 operators as,
\begin{eqnarray}
\label{mass5}
-{\mathcal{L}}_{\rm Yukawa}^{\nu} &=&      y_{ij}^\nu \bar{ \psi}_{L_i}^\ell   \tilde{\varphi}  \nu_{f_R} \left[  \dfrac{ \chi_i \chi_j (\text{or}~ \chi_i  \chi_j^\dagger)}{\Lambda_{\rm F}^2} \right] +  {\rm H.c.}. 
\end{eqnarray}
The Dirac mass matrix for neutrinos becomes,
\begin{equation}
\label{NM}
\M_{\D} = \dfrac{v}{\sqrt{2}}  
\begin{pmatrix}
y_{11}^\nu   \epsilon_1 \epsilon_7   &  y_{12}^\nu   \epsilon_4 \epsilon_7  & y_{13}^\nu  \epsilon_4  \epsilon_7 \\
0   & y_{22}^\nu  \epsilon_4  \epsilon_7 &  y_{23}^\nu  \epsilon_4  \epsilon_7 \\
0   &   y_{32}^\nu  \epsilon_5  \epsilon_7   &  y_{33}^\nu  \epsilon_5  \epsilon_7
\end{pmatrix}.
\end{equation}

The neutrino masses turn out to be,
\begin{eqnarray}
\label{mass11}
m_3  &\approx&  |y^\nu_{33}|  \epsilon_5 \epsilon_7 v/\sqrt{2}, 
m_2  \approx     |y^\nu_{22} - \dfrac{y_{23}^\nu  y_{32}^\nu}{y_{33}^\nu} |  \epsilon_4 \epsilon_7 v /\sqrt{2},
m_1  \approx  |y_{11}^\nu  |\,  \epsilon_1 \epsilon_7 v /\sqrt{2}, \quad \quad
\end {eqnarray}
and leptonic mixing angles are,
\begin{eqnarray}
\sin \theta_{12}^\ell  \approx  1 - 2 \sin \theta_{12}, 
\sin \theta_{23}^\ell  \approx  1 -  \sin \theta_{12}, 
\sin \theta_{13}^\ell   \approx \sin \theta_{12} - \frac{m_s}{m_c}.
\end{eqnarray}
For more details, see reference \cite{Abbas:2023dpf}.

\section{Dark-technicolour at low scale}
The DTC paradigm provides  an ultra-violet completion  to SHVM,  and thus provides a solution to the flavour problem of the SM.  Moreover,  it also has built-in dark-matter in the form of the flavonic dark matter.  The DTC model  is based on the $\mathcal{S} =  SU(3)_c \times SU(2)_L \times U(1)_Y \times SU(\rm{N}_{\rm{TC}}) \times SU(\rm{N}_{\rm{DTC}}) \times SU(\rm{N}_{\rm{F}})  $ symmetry, where   $\rm F$ stands for  a strong dynamics of vector-like fermions  \cite{Abbas:2020frs}.   The TC sector consists of the following  doublets  of fermions transforming under   $\mathcal{S} $ as \cite{Abbas:2020frs},
\begin{eqnarray}
{T_L}^i  &\equiv&   \begin{pmatrix}
T_i  \\
B_i
\end{pmatrix}_L:(1,2,0, \rm{N}_{ TC},1,1),  
T_{i, R} : (1,1,1, \rm{N}_{ TC},1,1), \\ \nonumber
 B_{i, R}  &:& (1,1,-1, \rm{N}_{ TC},1,1), 
\end{eqnarray}
where electric charges $+\frac{1}{2}$ for $T$ and $-\frac{1}{2}$ for $B$.  There are no anomalies involving the hypercharge since the hypercharges of the $T_R$ and $B_R$ are opposite. 

For the $\rm DTC$ dynamics we have,
\begin{eqnarray}
 D  &\equiv& C_{i,  L,R}: (1,1, 1,1, \rm{N}_{\rm DTC},1),~ S_{i, L,R} : (1,1,-1,1, \rm{N}_{\rm DTC},1), 
\end{eqnarray} 
where   electric charges $+\frac{1}{2}$ for $\mathcal C$ and $-\frac{1}{2}$ for $\mathcal S$.   

The $ SU(\rm N_{\rm F})$ dynamics has vector-like fermions transforming as,
\begin{eqnarray}
F_{L,R} &\equiv &U_{L,R}^i \equiv  (3,1,\dfrac{4}{3},1,1, \rm{N}_{\rm F}),
D_{L,R}^{i} \equiv   (3,1,-\dfrac{2}{3},1,1,\rm{N}_{\rm F}),  \\ \nonumber 
N_{L,R}^i &\equiv&   (1,1,0,1,1,\rm{N}_{\rm F}), 
E_{L,R}^{i} \equiv   (1,1,-2,1,1,\rm{N}_{\rm F}).
\end{eqnarray}

The following Lagrangian will produce the SM Higgs and the gauge singlet scalar fields $\chi_i$ of the SHVM,
\begin{eqnarray}
\mathcal{L} & =&  -\frac{1}{4}{\cal F}_{TC,\mu\nu}^{a} {\cal F}^{a\mu\nu}_{TC} -\frac{1}{4}{\cal F}_{DTC,\mu\nu}^{a} {\cal F}^{a\mu\nu}_{DTC} -\frac{1}{4}{\cal F}_{F,\mu\nu}^{a} {\cal F}^{a\mu\nu}_{F}  \\ \nonumber 
&& + i\bar{T}_L
\gamma^{\mu}D_{\mu}^{TC}T_L 
+ i\bar{T}_{i,R} \gamma^{\mu}D_{\mu}^{TC}T_{i,R}   +
i\bar{B}_{i,R} \gamma^{\mu}D_{\mu}^{TC}B_{i,R}, \\ \nonumber 
&& + i\bar{C}_{i,L}
\gamma^{\mu}D_{\mu}^{DTC}C_{i,L} 
+ i\bar{C}_{i,R} \gamma^{\mu}D_{\mu}^{DTC}C_{i,R}   + i\bar{S}_{i,L}
\gamma^{\mu}D_{\mu}^{DTC}S_{i,L} 
+ i\bar{S}_{i,R} \gamma^{\mu}D_{\mu}^{DTC}S_{i,R},  
\\ \nonumber 
&& + i\bar{U}_{i,L}
\gamma^{\mu}D_{\mu}^{F}U_{i,L} 
+ i\bar{U}_{i,R} \gamma^{\mu}D_{\mu}^{F}U_{i,R}   + i\bar{D}_{i,L}
\gamma^{\mu}D_{\mu}^FD_{i,L} 
+ i\bar{D}_{i,R} \gamma^{\mu}D_{\mu}^FD_{i,R}, 
\\ \nonumber 
&& + i\bar{N}_{i,L}
\gamma^{\mu}D_{\mu}^FN_{i,L} 
+ i\bar{N}_{i,R} \gamma^{\mu}D_{\mu}^FN_{i,R}   + i\bar{E}_{i,L}
\gamma^{\mu}D_{\mu}^FE_{i,L} 
+ i\bar{E}_{i,R} \gamma^{\mu}D_{\mu}^F E_{i,R},  
\end{eqnarray}
where the  field strength is ${\cal F}_{j,\mu\nu}^a =
\partial_{\mu}{\cal A}_{j, \nu}^a - \partial_{\nu}{\cal A}_{j, \mu}^a + g_{j} \epsilon^{abc} {\cal A}_{j,\mu}^b
{\cal A}_{j,\nu}^c,\ a,b,c=1,\ldots N^2-1$ for an $SU(\rm N)$ gauge group, and $j$ stands for TC, DTC, and F.
The covariant derivative for the  left-handed techniquarks is given as,
\begin{eqnarray}
D_{\mu}^{TC} T^a_L &=& \left(\delta^{ac}\partial_{\mu} + g_{TC}{\cal
A}_{TC, \mu}^b \epsilon^{abc} - i\frac{g}{2} \vec{W}_{\mu}\cdot
\vec{\tau}\delta^{ac} -i g'\frac{y}{2} B_{\mu} \delta^{ac}\right)
T_L^c, \ 
\end{eqnarray}
where ${\cal A}_{TC, \mu}^b$ is the TC gauge field, $\vec{W}_{\mu}$ and  $B_{\mu}$ are the SM gauge fields, $\tau^a$ are the Pauli matrices, and $\epsilon^{abc}$ is the fully antisymmetric tensor.

In this model, we have three axial $U(1)_A^{\rm TC, DTC,  F}$ symmetries.  They are broken to  a  discrete cyclic group as,  $ U(1)_A^{\rm TC, DTC, F} \rightarrow \mathcal{Z}_{2 \rm K_{\rm TC, DTC, F}}$ \cite{Harari:1981bs},  where $\rm K_{\rm TC, DTC, F} $ are number of  fundamental massless flavours of the TC, DTC, and F gauge dynamics in the $N$-dimensional representation of the gauge group $SU(\rm N)_{\rm TC, DTC, F}$.   This results in  a generic   $\mathcal{Z}_{\rm N} \times \mathcal{Z}_{\rm M} \times \mathcal{Z}_{\rm P}$ flavour symmetry, where $\rm N= 2 \rm K_{\rm TC}$, $\rm M= 2 \rm K_{\rm DTC}$ and $\rm P= 2 \rm K_{\rm F}$.   This symmetry is used to achieve the flavour structure of SHVM  \cite{Abbas:2023dpf}.

\begin{figure}[h]
	\centering
 \includegraphics[width=\linewidth]{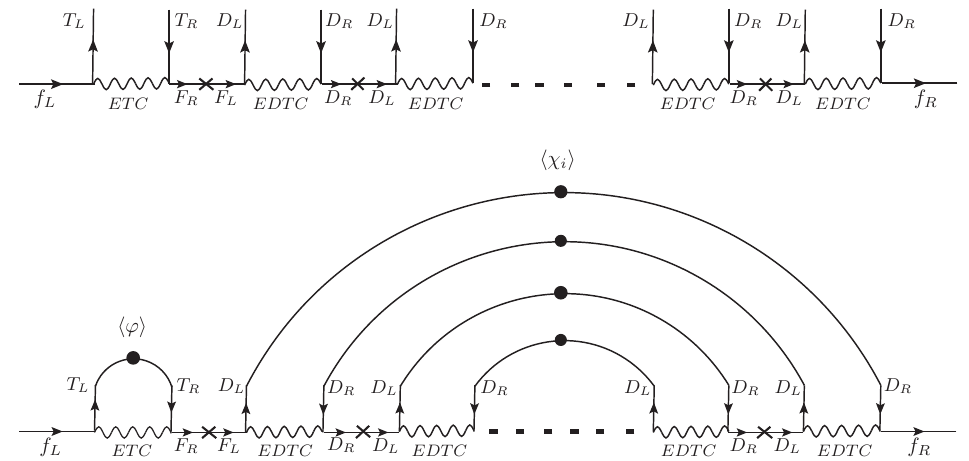}
    \caption{The Feynman diagrams for the masses of the quarks and charged leptons in the  dark-technicolour paradigm.  On the top, the generic interactions of the SM, TC, F and DTC fermions are shown.  In the bottom, we see the formations of the different condensates (circular blobs),    and the resulting mass of the SM fermions.}
 \label{fig1}	
 \end{figure}

In the DTC paradigm,  the masses of fermions of SHVM originate from the generic interactions given in figure \ref{fig1} on the top.  The formation of consdensates and resulting  mass of the SM fermion is shown below  in figure \ref{fig1}.   By deviating from reference  \cite{Abbas:2020frs}, we assume that there are two separate ETC for the TC and DTC gauge sectors.  This means that the TC fermions,  the left-handed SM fermions and the $F_R$ fermions are accommodated in an ETC symmetry.  On the other hand, the DTC fermions,  the right-handed SM singlet fermions and the $F_L$ fermions are accommodated in a dark-extended-TC (EDTC) symmetry.  Thus,  the $SU(\rm N)_{\rm F}$ symmetry plays the role of a connecting bridge between the TC and DTC sectors.   Due to this,  the mixing between the TC and the DTC dynamics is suppressed by the $1 / \Lambda_{\rm F}$ factor.  This is the reason that the discovered Higgs is behaving like the SM Higgs boson.   The mass of a charged fermion is now given by,
\bea
\label{TC_masses}
m_{f} & = & y_f  \frac{\Lambda_{\text{TC}}^{3}}{\Lambda_{\text{ETC}}^2}  \dfrac{1}{\Lambda_{\text{F}}} \frac{\Lambda_{\text{DTC}}^{n_i + 1}}{\Lambda_{\text{EDTC}}^{n_i}} \exp(n_i k),~
\eea
where $n_i = 2,4,6, \cdots 2 n $ is the number of fermions in a multi-fermion chiral condensate that plays the role of the VEV $ \langle \chi_i \rangle$ \cite{Abbas:2020frs}.  The multi-fermion condensate is parametrized as  \cite{Aoki:1983yy}, 
\be 
\label{VEV_h}
\langle  ( \bar{\psi}_L \psi_R )^n \rangle \sim \left(  \Lambda \exp(k \Delta \chi) \right)^{3n},
\ee
where $\Delta \chi$ denotes the chirality of an operator, $k$ is a constant, and $\Lambda$ shows the scale of the underlying gauge theory.   From this,  we can identify in equation \ref{TC_masses},
\bea
\epsilon_i =  \dfrac{1}{\Lambda_{\text{F}}} \frac{\Lambda_{\text{DTC}}^{n_i + 1}}{\Lambda_{\text{EDTC}}^{n_i}} \exp(n_i k).
\eea

\begin{figure}[h]
	\centering
 \includegraphics[width=\linewidth]{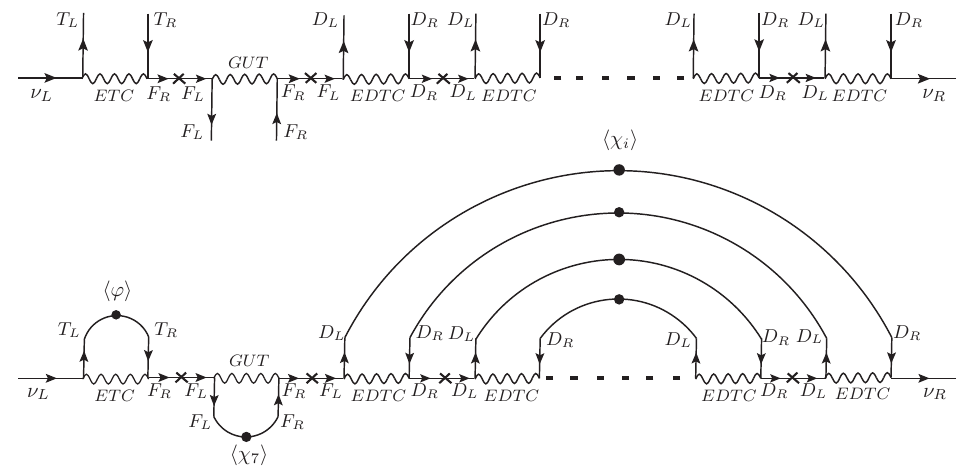}
    \caption{The Feynman diagrams for the masses of neutrinos in the dark-technicolour paradigm. On the top, there are generic interactions involving the SM, TC, DTC gauge sectors mediated by ETC, EDTC and GUT gauge bsons.  In the bottom, we show the generic Feynman diagram after the formation of the fermionic condensates. }
 \label{fig2}	
 \end{figure}
 
In  SHVM, neutrino masses originate from dimension-6 operators \cite{Abbas:2023dpf}, which are generated by the generic interactions given in figure \ref{fig2} on the top.  For generating these interactions, it is required that the TC and DTC sectors are unified in a GUT theory where the GUT gauge bosons mediate the interactions  between the $F_L$ and $F_R$ fermions.  Moreover, the chiral condensate $\langle \bar{F}_L F_R \rangle$ plays the role of the VEV $\langle \chi_7 \rangle$.  Thus, the mass of a neutrino is given by,
\bea
\label{TC_nmasses}
m_{\nu} & = & y_f  \frac{\Lambda_{\text{TC}}^{3}}{\Lambda_{\text{ETC}}^2}  \dfrac{1}{\Lambda_{\text{F}}} \frac{\Lambda_{\text{DTC}}^{n_i + 1}}{\Lambda_{\text{EDTC}}^{n_i}} \exp(n_i k)  \dfrac{1}{\Lambda_{\text{F}}} \frac{\Lambda_{\text{F}}^{3}}{\Lambda_{\text{GUT}}^{2}} \exp(2 k).
\eea
We now fit the masses and mixing-parameters of fermions to experimental data.  For the charged fermions, we use the masses given in reference \cite{Xing:2007fb} at 1 TeV. The quark-mixing angles are taken from reference \cite{pdg22}, and neutrino oscillation data for the normal-ordering are used from reference \cite{deSalas:2020pgw}.

Moreover, we assume that in a larger unified theory such as a  GUT, the couplings of the TC, F, and DTC sectors are unified as,
\be
\alpha_{\rm TC}(\rm M_{GUT}) = \alpha_{\rm F}(\rm M_{GUT}) = \alpha_{\rm DTC }( \rm M_{GUT}) =  \alpha_{3 }( \rm M_{GUT}).
\ee
This helps us to  use the renormalization group evolution to determine the scales $\Lambda_{\rm TC, DTC, F}$   through the following equations,
\be 
\label{rg1}
\frac{\Lambda_{\rm F} }{\Lambda_{\rm QCD} } = \exp\Bigl[   \dfrac{2\pi (\beta_0^{\rm F} - \beta_0^{\rm QCD})}{\beta_0^{\rm F} \beta_0^{\rm QCD} \alpha_{3 }( \rm M_{GUT})}\bigr],
\ee
and
\be 
\label{rg2}
\frac{\Lambda_{\rm TC} }{\Lambda_{\rm DTC} } = \exp\Bigl[   \dfrac{2\pi (\beta_0^{\rm TC} - \beta_0^{\rm DTC})}{\beta_0^{\rm TC} \beta_0^{\rm DTC} \alpha_{3 }( \rm M_{GUT})}\bigr],
\ee
where,
\be 
 \beta_0^{\rm QCD} = 11  - \frac{2}{3} n_f,~\beta_0^{\rm TC} = \frac{11 \rm{N}_{\rm T}}{3} - \frac{4}{3} \rm{N}_{\rm D}, ~ \beta_0^{\rm F, DTC } = \frac{11 \rm{N}_{\rm F, DTC}}{3} - \frac{2}{3} n_{\rm F, DTC},
\ee
where $\rm N_D$ are number of doublets and $n_{\rm f, TC, F, DTC}$ show the  number of massless flavours. 

We perform the fit after including equations \ref{rg1} and \ref{rg2}, assuming the simplest scenario where $\rm N_D = 1$ and $k=1$.  For $ \alpha_{3 }( \rm M_{GUT}) = 1/24.3$, $\Lambda_{\rm QCD}= 210 \text{ MeV}$ \cite{pdg22}, $ n_f = 12$ and    $v=246.22$ GeV, we predict,
 \begin{eqnarray}
 && \rm{N}_{\rm T} = 4, \rm{N}_{\rm F} =5, \rm{N}_{\rm DTC} = 6, n_{\rm F} = 22, n_{\rm DTC} = 12, \Lambda_{\rm GUT}= 1.18 \times 10^8 \text{ GeV}\nonumber \\ 
 &&    \Lambda_{\rm TC}= 190.546  \text{ GeV} , \Lambda_{\rm ETC}= 10^7 \text{ GeV}, \Lambda_{\rm DTC}= 328.713 \text{ GeV},  \nonumber \\ 
 &&\Lambda_{\rm EDTC}= 398.52 \text{ GeV}, \Lambda_{\rm F} = 2.19 \times 10^3 \text{ GeV}, \nonumber \\ 
 &&  \{n_1, n_4, n_5, n_2, n_6, n_3\}  = \{10, 16, 18, 22, 24, 28\}.
 \end{eqnarray}
The scan results for generation of the Yukawa couplings $y_{ij}^{u,d,\ell,\nu}= |y_{ij}^{u,d,\ell,\nu}| e^{i \phi_{ij}^{q,\ell,\nu}}$, in the range $|y_{ij}^{u,d,\ell, \nu}| \in [0.9,2 \pi ]$ and $ \phi_{ij}^{q,\ell,\nu} \in [0,2\pi]$ are, 
\begin{equation*}
y^u_{ij} = \begin{pmatrix}
-4.71 - 3.72 i & 0 & -1.02 + 0.08 i  \\
0 & -0.87 - 0.46 i &  -0.91 - 4.85 i \\
0 & 0.90 + 0.17 \times 10^{-3} i &  2.20 + 0.102 i
\end{pmatrix},  
\end{equation*}

\begin{equation*}
y^d_{ij} = \begin{pmatrix}
0.21 - 2.92 i & - 3.25 + 0.92 i & 3.14\, -0.06 i   \\
4.84\, + 2.50 i & - 2.82\, + 0.91 i &  4.55\, + 0.64 i \\
0.88\, + 0.19 i & 0.97+ 0.15 i &  - 0.61 - 0.66 i
\end{pmatrix},  
\end{equation*}

\begin{equation*}
y^\ell_{ij} = \begin{pmatrix}
6\, -0.2 \times 10^{-4} i & 1 & 1 \\
0  & -4.58\, + 1.85 i & 1 \\
0 & 0 & 3.31 + 1.65 \times 10^{-7} i
\end{pmatrix},
\end{equation*}

\begin{equation*}
y^\nu_{ij} = \begin{pmatrix}
- 0.25\, + 1.22 i & 1 & 1 \\
0 & - 1.18\, - 0.94 i & 0.42 + 1.30 i \\
0 & 0.03 + 1.32 i & 0.15\, - 0.92 i
\end{pmatrix}.
\end{equation*}

\section{Prediction of a  dark-Higgs decaying to $\gamma \gamma $}
The spectrum of the TC models  can be estimated  by scaling up the QCD  in 't Hooft large-$N$ limit  \cite{tHooft:1973alw,Witten:1979kh}. 
This predicts a heavier Higgs boson, approximately $1.5$ TeV.  However, the Higgs boson mass is found to be 125 GeV by the LHC. We solve this problem by assuming that the DTC spectrum  in the DTC paradigm is a scaled-up version of the TC spectrum.  Moreover, we have already seen that the DTC spectrum consists of multi-fermion hierarchical bound states, therefore, the TC spectrum will also follow the same pattern.  For the large-$N$ limit of such QCD-like theories, one needs to modify the simple scaling rules used in earlier TC models.  We use the scaling  discussed in reference \cite{Sannino:2008ha}. 

The TC scalar bound state, which is the discovered Higgs boson, is related to its dark counterpart, referred  to as  ``dark-Higgs" (DH), as  \cite{Sannino:2008ha},
\bea
m_{\rm H } = \frac{\sqrt{2} v }{F_{\rm DTC} \sqrt{\rm{N}_{\rm D}}} \left(  \frac{\rm{N}_{\rm TC}}{\rm{N}_{\rm DTC}}  \right)^{\frac{p-1}{2}} m_{\rm DH },
\eea 
where $p \geq 0$. 

Using the  fitting results and $p=0$, we predict the dark-Higgs mass to be approximately $ m_{\rm DH } = 95.4$ GeV for $F_{\rm DTC} = 324.5$ GeV.  This is similar to the excess reported by the CMS and ATLAS  in the diphoton invariant-mass distribution at $m_{\gamma\gamma} = 95.4$ \cite{CMS:2023yay,ATLAS1,ATLAS2}.

There are no modes of the dark-Higgs where it decays to the SM matter fields.  Moreover, it decays of the type $H_D \rightarrow f_R F_L$ are forbidden by the kinematics.  Thus, the only  decay mode is provided by the electromagnetic interactions as $H_D \rightarrow \gamma \gamma$, which is a smoking-gun signature of the DTC paradigm.  

For TC and DTC vector mesons, we assume that they are multi-fermion states, and  their masses can be estimated by, 
\bea
m_{\rm T_{\rho}} = \frac{\sqrt{2} v }{F_{\rm DTC} \sqrt{\rm{N}_{\rm D}}} \left(  \frac{\rm {N}_{\rm TC}}{\rm{N}_{\rm DTC}}  \right)^{\frac{p-1}{2}} m_{\rm DTC_{\rho}}.
\eea 

\begin{figure}[h]
	\centering
 \includegraphics[width=0.55\linewidth]{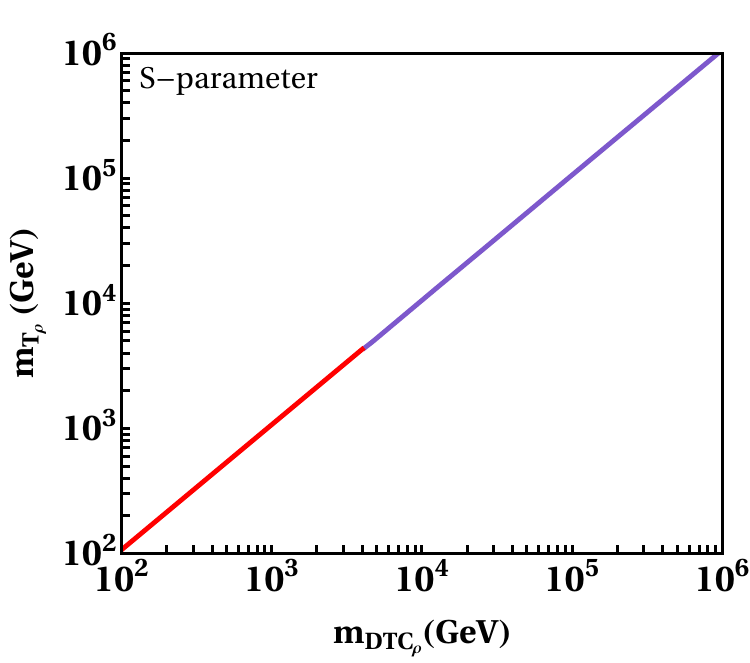}
    \caption{The allowed mass ranges of the TC and DTC vector mesons by oblique parameter $S$ are shown with the blue-coloured line. The red-coloured line represents their mass ranges, which are excluded by the $S$-parameter.}
 \label{fig3}	
 \end{figure}
 
The mass of the TC meson $m_{\rm T_{\rho}}$ can be constrained using the oblique parameters $S$ and $T$ \cite{Peskin:1990zt,Peskin:1991sw}.  We use the following values of the  $S$ and $T$ parameters  \cite{pdg22},
\be  
S=-0.02\pm 0.1,~~ T= 0.03\pm 0.12.
\ee
These parameters at  next-to-leading-order  read as \cite{Pich:2013fea},
\begin{align}
S &=4\pi F_\Pi^2 \left[ \frac{1}{M_V^2}+\frac{1}{M_A^2} \right] + \frac{1}{12 \pi} \left[\left( 1-\frac{M_V^4}{M_A^4}\right) \left(   \log \frac{M_V^2}{M_{\rm Higgs}^2} -\frac{11}{6}  \right) + \left( \frac{M_V^2}{M_A^2}-\frac{M_V^4}{M_A^4}\right)  \log \frac{M_A^2}{M_V^2}  \right], \\ \nonumber
T & = \frac{3}{16 \pi \cos^2\theta_W} \left[ (1-\kappa_W^2)      \left(  1+ \log\frac{M_{\rm Higgs}^2}{M_V^2}  - \kappa_W^2   \log\frac{M_V^2}{M_A^2}     \right)                \right],
\label{s1}
\end{align}
where the $\kappa_W = \frac{M_V^2}{M_A^2}$.

In figure \ref{fig3}, we show the bounds derived on the masses of the TC meson $m_{\rm T_{\rho}}$ and the DTC meson $m_{\rm DTC_{\rho}} $ using the most recent value of the oblique parameters $S$ and $T$  assuming  $\kappa_W = 1$ and $p=1$.  We observe that the masses of the mesons $m_{\rm T_{\rho}}$ and $m_{\rm DTC_{\rho}} $  are constrained from below to be $m_{\rm T_{\rho}} \geq 4.37 \times 10^3$ GeV and $m_{\rm DTC_{\rho}} \geq 4.07 \times 10^3 
 $ GeV.

\section{Summary}
TC models were invented to provide a natural mechanism for the electroweak symmetry breaking.  However, QCD-like TC models failed to produce the fermionic mass spectrum of the SM.  In this work, we have discussed a dark-technicolour model with  a scale close to the eletroweak-scale, and capable of solving the flavour problem by producing the flavour structure of the SHVM.  A defining signature of this  scenario is the decay $H_D \rightarrow \gamma \gamma$, where the mass of the dark-Higgs is approximately around 95.4 GeV in the simplest scenario.  This may be probed by the LHC in the future high-luminosity phase.

\section*{Acknowledgement}
This work is supported by the  Council of Science and Technology,  Govt. of Uttar Pradesh,  India through the  project ``   A new paradigm for flavour problem "  no. CST/D-1301,  and Science and Engineering Research Board,  Department of Science and Technology, Government of India through the project ``Higgs Physics within and beyond the Standard Model" no. CRG/2022/003237.  NS acknowledges the support from the  INSPIRE fellowship by the Department of Science and Technology, Government of India.

%%use \balance somewhere in the left column of the last page to balance the two columns in the end page

%%References section

\end{document}